# An apparent paradox in hydrostatic

E. N. Miranda
Área de Ciencias Exactas
CCT- CONICET-Mendoza
5500 –Mendoza, Argentina
and
Departamento de Física
Universidad Nacional de San Luis
5700 - San Luis, Argentina

**Abstract:**
It is shown that for a liquid in any connected vessels system, it is not possible to fulfill simultaneously Pascal´s principle, mass conservation, and energy conservation. The viscosity has to necessarily be taken into account to understand the system change.

There are subjects taught every year in General Physics courses that seem to be simple and even dull. Hydrostatic is one of them; however in this note it will be shown that a well-known system –the connected vessels- may surprise us. In this note it will be shown that Pascal´s principle and mass conservation imply necessarily that energy cannot be conserved in connected vessels under certain circumstances.

The initial experimental situation is shown in Figure 1. There are three connected vessels filled with a liquid of density $\rho$. The cross sections of the vessels are $A_1$, $A_2$ and $A_3$. Initially the fluid reaches the same height $h_0$ in the three branches. It is assumed that plates of negligible mass are on the liquid surfaces. Additionally a mass $m$, at a height $h_0$ is also taken into account.

The final situation is shown in Figure 2. The mass is on the plate of area $A_1$ and the liquid heights are now different in each branch: $h_1$, $h_2$ and $h_3$. Our aim is to evaluate the values of $h_1$, $h_2$ and $h_3$ and see what happens with energy conservation. For the particular configuration considered in this letter, it will come out that $h_2 = h_3$ but they may be not equal for a different design of the vessels.

The pressures in $a$, $b$ and $c$ should be equal (Pascal´s principle); in mathematical terms:

$$
\begin{aligned}
p_a &= p_b \\
&= p_c \\
\frac{mg}{A_1} + \rho g h_1 &= \rho g h_2 \\
&= \rho g h_3.
\end{aligned}
\tag{1}
$$

The mass is conserved; therefore the liquid lost in vessel 1 should be equal to the liquid gains in vessels 2 and 3:

$$
A_1(h_0 - h_1) = A_2(h_2 - h_0) + A_3(h_3 - h_0). \tag{2}
$$

This author thought that the energy should also be conserved but he was wrong. One should remember that the potential energy of liquid column of height $h$ is $\tfrac{1}{2} g\, \rho A h^2$. Then, the initial $E_i$ and final $E_f$ energies of our experiment are:

$$
\begin{aligned}
E_i &= mgh_0 + \tfrac{1}{2}\rho g A_1 h_0^2 + \tfrac{1}{2}\rho g A_2 h_0^2 + \tfrac{1}{2}\rho g A_2 h_0^2 \\
E_f &= mgh_1 + \tfrac{1}{2}\rho g A_1 h_1^2 + \tfrac{1}{2}\rho g A_2 h_2^2 + \tfrac{1}{2}\rho g A_2 h_3^2.
\end{aligned}
\tag{3}
$$

The last two lines of Eq. (1) and Eq. (2) are a linear system of equations with a unique solution for the values of $h_1$, $h_2$ and $h_3$. Equation (3) has not been used. Let us see what happens with the energy, and it should be remarked that nothing has been said about the liquid viscosity.

The energy difference $\Delta E$ is:

$$\Delta E = E_f - E_i \\
= mg(h_1 - h_0) + \tfrac{1}{2}\rho g A_1 (h_1^2 - h_0^2) + \tfrac{1}{2}\rho g A_1 (h_2^2 - h_0^2) + \tfrac{1}{2}\rho g A_1 (h_2^2 - h_0^2). \quad (4)$$

Equation (4) can be rewritten as:

$$\Delta E = (h_1 - h_0) g \{ m + \tfrac{1}{2}\rho A_1 (h_1 + h_0) + \tfrac{1}{2}\rho A_2 (h_2 + h_0) + \tfrac{1}{2}\rho A_3 (h_3 + h_0) \}. \quad (4')$$

The expression between brackets is always positive while $(h_1-h_0)$ is always negative. For that reason, it is $\Delta E < 0$ for any geometrical parameters of the vessels or physical properties of the liquid. Where has the energy gone?

The answer to that question is a tacit assumption made about the liquid viscosity. When the mass $m$ is put on the liquid surface in vessel 1, the system becomes unstable and it starts to oscillate. To reach the new equilibrium situation shown in Figure 2, the oscillations have to fade away and this implies the liquid is viscous. Some energy has been dissipated due to that property of the fluid.

This hydrostatic paradox is analogue to that of a couple of charged capacitors that are suddenly connected and some energy is "missed" [1, 2]. An analysis shows that the energy is lost due to Joule heat in the connecting wires. The resistance of the wire cannot be zero if the system attains a new static configuration.

In summary, an inattentive analysis of the situation displayed in Figures 1 and 2 lead us to consider that three principles should hold simultaneously: Pascal´s, mass conservation and energy conservation. This assumption is wrong. The energy is not conserved because some dissipation is needed to go from the initial to the final states. It has been tacitly supposed the liquid is viscous and energy is not constant. A not difficult exercise generalizes this conclusion for any disposition of the vessels.

**Acknowledgement**: the author thanks the National Research Council of Argentina (CONICET) for financial support.

# Figure captions:

## Figure 1:
This is the initial state of the analyzed system. There are three connected vessels open to the atmosphere and the liquid heights $h_0$ are the same for all branches. The liquid surfaces are cover with plates of negligible weight shown with dash lines in the figure. A mass $m$ that is initially at the same height $h_0$ is also taken into account. We are interested in the evaluation of the energy change when the system goes from the situation shown in this figure to that of Figure 2.

## Figure 2:
The mass $m$ has been put on the plate of the first branch, i.e. there is an extra pressure $mg/A_1$ on the liquid of this vessel. The heights in the three vessel change to $h_1$, $h_2$ and $h_3$. Because of Pascal´s principle the pressure in points $a$, $b$ and $c$ has to be the same. For this particular configuration it results that $h_2 = h_3$, but these heights has been plotted as if they were different to take into account other configurations of three vessels. Pascal´s principle plus mass conservation determine univocally the heights values. However, it comes out that energy is not conserved.

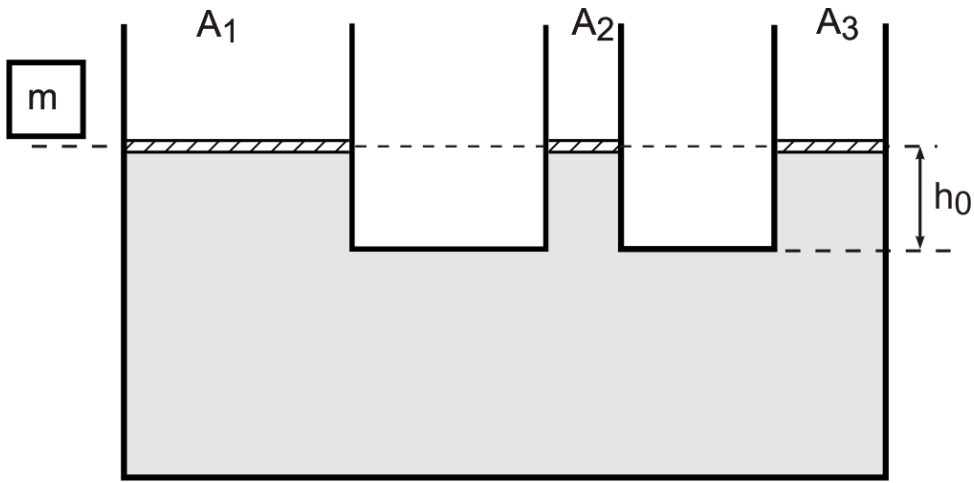

**Figure 1**

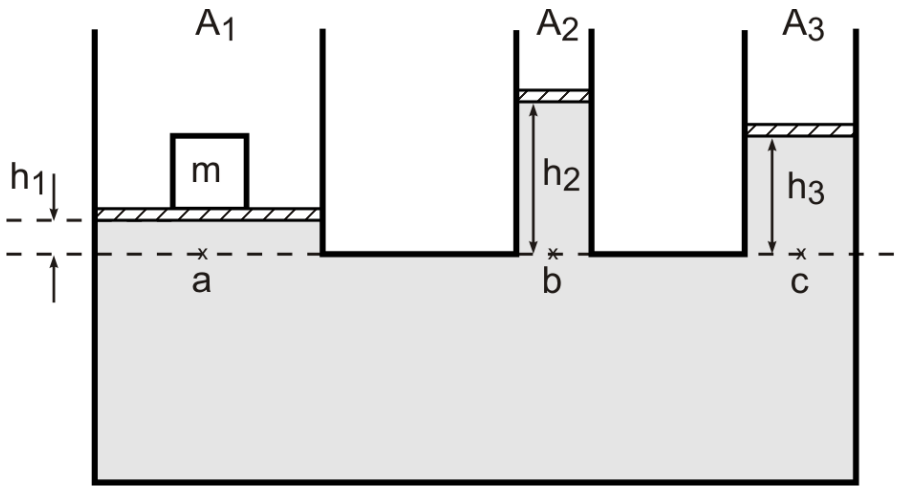

**Figure 2**

## Corrigendum

**An apparent paradox in hydrostatics**
E N Miranda 2009 *Eur. J. Phys.* **30** L55–57

In equation (3) the last term depends on $A_3$ and not on $A_2$. Therefore, the initial and final energies are:

$$E_i = mgh_0 + \tfrac{1}{2}\rho g A_1 h_0^2 + \tfrac{1}{2}\rho g A_2 h_0^2 + \tfrac{1}{2}\rho g A_3 h_0^2$$
$$E_f = mgh_1 + \tfrac{1}{2}\rho g A_1 h_1^2 + \tfrac{1}{2}\rho g A_2 h_2^2 + \tfrac{1}{2}\rho g A_3 h_3^2. \qquad (3)$$

In equation (4), the area $A_1$ that appears in the last two terms should be changed to $A_2$ and $A_3$ respectively. The correct equation reads:

$$\Delta E = E_f - E_i$$
$$= mg(h_1 - h_0) + \tfrac{1}{2}\rho g A_1(h_1^2 - h_0^2) + \tfrac{1}{2}\rho g A_2(h_2^2 - h_0^2) + \tfrac{1}{2}\rho g A_3(h_2^2 - h_0^2). \qquad (4)$$

There is a better approach for evaluating $\Delta E$. Solving the linear equations system written in equation (1), one gets the values of $h_1$ and $h_2$ in terms of the geometrical and physical parameters of the experiment. Replacing those values in (4), the energy difference can be written as:

$$\Delta E = -\frac{1}{2}\frac{m^2 g}{\rho}\frac{A_2 + A_3}{A_1(A_1 + A_2 + A_3)}. \qquad (4')$$

There is always energy lost for any value of the system parameters, i.e., the conclusion of the original paper is correct.